\bigskip

\centerline{\bf NONCOMMUTATIVE  GEOMETRY, NEGATIVE PROBABILITIES          }
\centerline{\bf  AND  CANTORIAN-FRACTAL  SPACETIME  } 

\bigskip

\centerline{ Carlos Castro} 
\centerline{ Center for Theoretical Studies of Physical Systems} 
\centerline{ Clark Atlanta University} 
\centerline{ Atlanta, Georgia 30314 } 
\smallskip

\centerline{ July, 2000} 

\bigskip

\centerline{\bf ABSTRACT} 
\bigskip

A straightforward explanation of the Young's two-slit experiment of a quantum particle is obtained 
within the framework of the Noncommutative Geometric associated with El Naschie's  Cantorian-Fractal 
transfinite Spacetime continuum.

\bigskip

\centerline{ \bf 1. Introduction} 

\bigskip

One of the most paradoxical conclusions of QM is related to an indivisible quantum particle traversing the 
Young's two-slit experiment; i.e the coexistence of a point particle at two separate spatial locations 
and the origins of the wave-particle duality in QM [1] . It was argued in [1] that provided spacetime is 
effectively a four-dimensional random manifold and the path of a quantum particle is fractal then any question about the exact spatial location of a microscopic point is fundamentable undecidable due to the inherent 
uncertainty and fuzziness of the geometrical structure of such space [1]. This uncertainty is caused by 
an intrinsic indistinguishability between $intersections$  and $unions$ within the subsets of the 
random effectively four-dim manifold, ${\cal E}^{(4)}_c $ . Such space is just one representative 
element of the 
infinite-dimensional Cantorian-Fractal spacetime transfinite continuum,   ${\cal E}^{(\infty)}_c $ . 
The latter continuum is just a representative of von Neumann's Noncommutative Geometry. 
Therefore a  `` point `` within   ${\cal E}^{(\infty)}_c $ can in a sense occupy 
$two$ different locations at the same time. 

El Naschie considered from the start the backbone random Cantor set $S^{ (0)}_c$ 
, whose dimensionality with probability {\bf one}  
equals  the Golden Mean, $\phi = (\sqrt 5 - 1)/2, $  as was shown by the Mauldin-Williams theorem  [2] 
and extended to all of ${\cal E}^{(\infty)}_c $ by [3].  
In essence, these randomly constructed spaces are related to the random homeomorphisms  which  
furnish a natural probability measure on the spaces of $all$ the probability distributions. 
For a crucial role that the `` probability on the space of all probability distributions ``  
has in the construction of Bell's inequalities and in von Mises frecuency approach to probablity theory, 
versus the standard 
Kolmogorov approach, see [4].    

The dimension of the $dual$ or complementary set to  $S^{ (0)}_c$, relative to the normal set given by the 
space  ${\cal E}^{(1)}_c    $, of dimension $1$, is : $ {\tilde d}^{ (0) }_c = 1 - {d}^{ (0) }_c = 1 - \phi = \phi^2$. 
Using the well-known  relation between the Hausdorff dimension of a fractal path and the Hurst exponent $H$ one finds [1] : 

$$d_{path} = {1\over H} = d^{ (2) }_c = {1\over \phi} = 1+ \phi. ~~~{\tilde d_{path}} = {\tilde d}^{ (2) }_c
=  { 1 \over 1 - \phi} = 
{ 1\over \phi^2} = (1+\phi)^2. \eqno (1)$$

El Naschie finally showed that if, and only if, there is an equivalence between $unions$ and 
$intersections$ in the concerned space then we must have : 

$$d_{critical} = d^{ (2) }_c + {\tilde d}^{ (2) }_c = {1\over \phi} +  { 1\over \phi^2} = 
{ \phi ( 1 +\phi )\over \phi^3}= 
{ 1\over \phi^3} =   { 1\over \phi}  { 1\over \phi^2} = d^{ (2) }_c {\tilde d}^{ (2) }_c = 4+\phi^3.         ~~~
1+\phi = {1\over \phi}  . \eqno (2)$$

Where $ 4+\phi^3 = d_{critical}= dim~{\cal E}^{(4)}_c = 4.236067...$. 
The critical dimension coincides exactly with the 
Hausdorff dimension of the random set ${\cal E}^{(4)}_c$ which is embedded  
densely  onto a smooth set of topological dimension equal to four. 
The backbone random set of dimension ${d}^{ (0) }_c = \phi$, a randomly constructed Cantor set, is embedded  densely onto a set of topological dimension zero : a `` point `` .  
This justifies the notation : $ d^{ (n) }_c $ is the Hausdorff dimension of a randomly constructed 
space embedded onto  a smooth manifold of integer topological dimension equal to $n$.  

In the next section we shall provide a probabilistic argument to El Naschie's  results by firstly  
following in detail the particular Peano-Hilbert path traversing the two-slits in a fashion to be 
described below. 
This provides a very natural explanation of $negative$  probabilities within the framework of 
Noncommutative Geometry. And furthermore why there is an asymmetry of the randomly constructed  Cantor Set 
${\cal E}^{(4)}_c$ used by El Naschie in the Young's two-slit experiment. 

Negative dimensions and Negative Entropies [5] were of crucial importance 
in the explicit numerical proof why the average spacetime  dimension, over an 
infinity of dimensions ranging from $ -2 $ to $ \infty$,  is of the order of $4+ \phi^3$  [6]. 
This lends further creedence that there is must be a deeper  resaon why we live in four domensions 
than the one provided by the compactification schemes of string, $M$ theory : these attempts 
are no explanation as to why we live in four dimensions.    
For a discussion of the history of negative probabilities, and for that matter complex probability in QM,  
see the report article by Muckenheim [7]. For their role in $p$-Adic Quantum Mechanics  see [8] . 
A detailed  discussion of why there is {\bf no} {\bf EPR} paradoxes within the framework 
of the New Relativity Theory [9], linked to Cantorian-Fractal spacetimes , see [10] .

\bigskip

\centerline{ \bf 2. Negative Probabilities and Noncommutative Geometry } 

\bigskip

The probabilistic extension of El Naschie's dimensional arguments 
to explain the Young's two-slit experiment within the framework of Cantorian-Fractal spacetime requires choosing the Peano-Hilbert path from the source to the detector traversing the two slits in a particular fashion. 
The source of particles ( say and electron  gun) is located far to the left  of the partition  
with the two slits $B,A$. The slit $B$ lies above the slit $A$. 
The detector lies far to the right of the partition.  Region {\bf I} is the one to the left of the partition. 
Region { \bf II} is to the right.

Imagine the electron's path ( from the souce to the partition) impinging on the upper portion of slit $B$ 
at an angle $\alpha$ ( which for simplicity we may take to be $45$ degrees ) . 
Once it crosses slit $B$ it experiences a zig-zag typical of a Peano-Hilbert geodesic fractal path 
in a $clockwise$ fashion : the zig-zagging in region {\bf II}, begins at point $1$ with a 
sharp $90^o$ turn to the left reaching the vertex $2$, where another $90^o$ turn to the right takes the electron to the vertex $3$. Another  two consecutive $90^o$ turns will take the particle from $3$ back to the 
beginning $5$, where the curve $12345$ almost `` closes `` at points $1,5$ inside the slit $B$.

To sum up : The traversal of the first iteration of the Peano-Hilbert geodesic curve in region {\bf II} 
to the right of the slit $B$ is performed $clockwise$.

Once back in slit $B$ the path zig-zags towards the point $6$ in region {\bf I} in a south-west 
bound fashion,  at an angle 
perpendicular to the initial impinging direction to the partition ( south-east).  
At vertex $6$, in region {\bf I}, it zig-zags in a $counterclockwise$ fashion beginning the second iteration   of a Peano-Hilbert fractal geodesic path moving towards the lower edge of slit $A$. It crosses the lower edge of 
slit $A$ and reaches vertex $7$ when it performs a  $90^o$ turn to the left reaching the vertex $8$. 
Another counterclockwise turn from $8$ 
back to the lower edge of slit $B$. It crosses slit $B$ {\bf again}  at vertex $9$ towards region {\bf I}. 
A  futher $90^o$ turn at vertex $9$ , south-west bound towards vertex $6$, and another $90^o$ turn,  to cross finally the upper edge of the slit $A$ ,  
reaching the detector in region {\bf II} at an angle $\alpha = 45^o$.  

To sum up : The traversal of the second iteration of the Peano-Hilbert geodesic curve in regions {\bf I} and {\bf II} surrounding slits $A$ and $B$  is perfomed $counterclockwise$.

Notice the crucial difference between the two Peano-Hilbert geodesic paths. 
In the first case it exists to the right of the slit $B$ while in the second case it exists in both regions {\bf I, II} of the partition and it winds around both slits $A, B$. This essential difference will account for the numerical results that follow.  

We will assign probablity  magnitudes to the regions at $A, B$ respectively :

$$ |p_A |= |p(A)|= {1\over d_A} = \phi. ~~~|p_B| = |p(B)|=  {1 \over d_B } = \phi^2. \eqno (3)$$

The physical explanation of this goes as follows : 
The region around slit $B$ is comprised of $3$ lines. One line is right-flowing and two lines are left 
flowing. 
The region around slit $A$ is comprised of only two right-flowing lines.  
It is natural to assign a $higher$ dimensionality to the region around slit $B$ than the one around $A$.  
The right flowing line in slit $B$ could correspond to an $upper$ bridge connecting regions {\bf I, II}. 
And the two left-moving lines flow along a $lower$  bridge connecting regions {\bf II, I} ( like a knot). 
In slit $A$ there is only one bridge ( where the two right-moving lines flow from {\bf I} to {\bf II} ). 

Furthermore,  in slit $B$ the net flux is $negative$ while in slit $A$ is $positive$. 
In slit $B$ there are two negative units of flux ( left moving) and one positive unit of flux ( right moving) giving a net flux of $ 1-2= - 1 $ ( left moving). In region $A$ there are two positive units of 
right moving flux so net flux is $2$ ( positive). 

For these reasons  we will assign 
$different$  magnitudes and signs to $p_B$ w.r.t $p_A$ :

$$ p_A = p(A)= {1\over d_A} = \phi. ~~~p_B = p(B)= - {1 \over d_B } = - \phi^2. \eqno (4)$$

The Noncommutative nature of Cantorian-Fractal spacetime [1] is expressed explicitly as :

$$ p ( A \wedge B ) = - p ( B \wedge A ) =  (\phi) ( - \phi^2)       =   - \phi^3. \eqno (5) $$

Using the standard definitions of conditional probabilities  :

$$ p(A|B) =  { p ( A \wedge B ) \over p(B) }    = { (- \phi^3) \over  ( -\phi^2)  }= \phi . \eqno (6a) $$
$$ p(B|A) =  { p ( B \wedge A ) \over p(A) }  = { (  \phi^3) \over  (  \phi )  }= \phi^2 . \eqno (6b) $$

$ p(A|B)$ is the conditional probability for $A$ provided $B$ {\bf has}  occured. 
$ p(B|A)$ is the conditional probability for $B$ provided $A$ {\bf has}  occured. 
$ p(A|B)$ is associated with the first iteration of the Peano-Hilbert geodesics . It is the probability that 
the Peano-Hilbert geodesic which left slit $B$ will zig-zag around point $6$ and enter the slit $A$ below. 
And vice versa,  $ p(B|A)$ is associated with the probability that the second iteration of the 
Peano-Hilbert geodesic, around slit $A$,  will wind around slit $B$ after having gone through slit $A$.

However, $p(A\wedge B) = - p( B\wedge A ) = - \phi^3  $ is the joint probability associated 
with both events $A, B$ to 
take place and coexist {\bf simultaneously}. This is possible in a cantorian-Fractal spacetime which 
is in essence a `` Noncommutative Pointless'' geometry.   
The Noncommutative nature of the Cantorian-Fractal spacetime ${\cal E}^{ ( \infty) }_c$ is the one responsible for the fact that  $p(A\wedge B) = -  p(B\wedge A) $.

To check that negative probabilities do  {\bf not} violate $unitarity$ we must verfy that the sums of all probablities does not exceed unity. In fact the net sum equals {\bf unity}  exactly ! :

$$p(A)+ p(B)+ p ( A \wedge B ) + p(A|B) + p(B|A) =  [  \phi - \phi^2 - \phi^3  ] + [ \phi +   \phi^2 ] = 1. \eqno (7a) $$

Since 

$$ [  \phi - \phi^2 - \phi^3  ] = \phi - \phi^2 ( 1+ \phi ) = \phi - {\phi^2 \over \phi}  = 0. ~~~
[ \phi +   \phi^2 ] = \phi ( 1 + \phi) = 1. \eqno (7b)$$

Is this all nothing but a numerical coincidence or design that we live in $4+\phi^3$ [6] ? 
And that $4+ \phi^3$ is also the average dimension  of the Cantorian-Fractal spacetime ${\cal E}^{(\infty)}_c$ :  $ < dim~ {\cal E}^{(i)}_c > = dim~ {\cal E}^{(4)}_c = 4+\phi^3 = 4.236067...$ [1] ? .Where 
$ i$  ranges from $-\infty$ to $\infty$.   

\bigskip

Concluding, the Noncommutative nature of Cantorian-Fractal spacetime [1] combined with the notion of 
the inherent $negative$  probability  assigned to a Peano-Hilbert `` loop ``  around the region of slit $B$, 
because its orientation is $opposite$  to the Peano-Hilbert `` loop''  located in the region around slit $A$, 
explains in a very straightforward ( but {\bf non-classical})  fashion the wave-particle duality of an indivisible quantum-particle. The Noncommutative Cantorian-Fractal spacetime is {\bf not } a classical space. 
This supports further the interpretation of the Schroedinger equation by Nagasawa as $two~dual $ diffusion 
equations [5] , one forward and the other backwards in time , 
which also agrees with Nottale's [5] fractal geodesics behaviour of the Young's two-slit experiment and 
El Naschie complex time  and G. Ord's spiral gravity models [1].

\bigskip

\centerline{\bf Acknowledgements} 

We are indebted to M.S. El Naschie for numerous conversations which led to this work and for his 
hospitality in Garmisch, Germany and Laggo Maggiore, Switzerland. 
We thank also A. Granik and D. Chakalov for many discussions and to A. Schoeller and family for a 
warm Austrian hospitality.

\centerline{\bf REFERENCES} 

1- M. S. El Naschie : `` On the Uncertainty of Cantorian Geometry and the Two-slit Experiment `` 

Jour. Chaos, Solitons and Fractals {\bf 9} (3) (1998) 517-529.
 
M. S. El Naschie :'' Remarks on Superstrings, Fractal Gravity, Nagaswa's Diffusion and Cantorian Spacetime `` 

Jour. Chaos, Solitons and Fractals {\bf 8} (11) (1997) 1873.

M. S. El Naschie :'' Young's double-slit experiment , Heisenberg uncertainty principle and 
correlations in Cantorian Spacetime `` . 
In Quantum Mechanics, Diffusion and Chaotic Fractals, eds M.S. El Naschie, O.E. Rossler 
and I. Prigogine. Elsevier, Oxford, 1995.

2-. D. Mauldin , S. C Williams : `` Random Recursive Constructions. 

Transactions of the American Mathematical Society, {\bf 295} ( 1986) 325-346. 

3-M. S. El Naschie : `` Nuclear spacetime theories, superstrings, monster group and applications ``  

Jour. Chaos, Solitons and Fractals {\bf 10} (2-3) (1999) 567-580.

G. Ord : `` The spiral gravity models ``. Jour. Chaos, Solitons and Fractals {\bf 10} (2-3) (1999) 567-580.

4-A. Khrennikov : `` Einstein, Bell, von Mises and Kolgomorov....'' quant-ph/0006016.

5-M. Conrad : Special issue of Chaos, Solitons and Fractals {\bf 8} (5) (1997) on : 
`` Nonlinear Dynamics, General Relativity, 

and the Quantum ( Uneven Flow of Fractal Time ).'' eds by G. Ord, M. Conrad, O. Rossler, and M. S El Naschie.  

6- C. Castro, A. Granik and M.S El Naschie : `` Why we live in $3+1$ dimesnions `` hep-th/0004152 v4.

7-W. Muckenheim : Phys. Reports {\bf 133} (1986) 338-401

8- A. Khrennikov : `` $p$-Adic valued distributions in Mathematical Physics 

`` Kluwer Acad. Publishers , Dordrecht (1994)

V. Vladimorov, I. Volovich and E. Zelenov : `` p-Adics Analysis in Mathematical Physics `` World Scientific .

9- C. Castro : `` Hints of a New Relativity principle from $p$-brane Quantum Mechanics `` 

to appear in J. Chaos, Solitons and Fractals.

C. Castro : `` The String Uncertainty Relations follow from the New Relativity Principle `` 

to appear in Foundations of Physics.

10- C. Castro, A. Granik : `` How the New Scale Relativity resolves some Quantum paradoxes `` . 

to appear in J. Chaos, Solitons and Fractals

11- L.Nottale : `` The Scale Relativity Program ``  J. Chaos, Solitons and Fractals {\bf 10} ( 2-3) (1999).  

`` La Relativite dans Tous ses Etats `` Hachette eds. Paris.

\bye